\documentclass[draft,tightenlines,nofootinbib,preprint,aps,eqsecnum,amsmath,amssymb]{revtex4}

\newcommand{\beq}{\begin{equation}}
\newcommand{\eeq}{\end{equation}}
\newcommand{\bea}{\begin{eqnarray}}
\newcommand{\eea}{\end{eqnarray}}

\newcommand{\sgn}{\epsilon}

\begin{document}

\title{Towards Relativistic Atomic Physics and Post-Minkowskian Gravitational Waves}

\medskip

\author{Luca Lusanna}

\affiliation{ Sezione INFN di Firenze and ACES Topical Team of ESA\\ Polo Scientifico\\ Via Sansone 1\\
50019 Sesto Fiorentino (FI), Italy\\ Phone: 0039-055-4572334\\
FAX: 0039-055-4572364\\ E-mail: lusanna@fi.infn.it}

\begin{abstract}

A review is given of the  formulation of relativistic atomic theory,
in which there is an explicit realization of the Poincare'
generators, both in the inertial and in the non-inertial rest-frame
instant form of dynamics in Minkowski space-time. This implies the
need to solve the problem of the relativistic center of mass of an
isolated system and to describe the transitions from different
conventions for clock synchronization, namely for the
identifications of instantaneous 3-spaces, as gauge transformations.
These problems, stemming from the Lorentz signature of space-time,
are a source of non-locality, which induces a spatial
non-separability in relativistic quantum mechanics, with
implications for relativistic entanglement.

Then the classical system of charged particles plus the
electro-magnetic field is studied in the framework of ADM canonical
tetrad gravity in asymptotically Minkowskian space-times admitting
the ADM Poincare' group at spatial infinity, which allows to get the
general relativistic extension of the non-inertial rest frames of
special relativity. The use of the York canonical basis allows to
disentangle the tidal degrees of freedom of the gravitational field
from the inertial ones. The York time is the inertial gauge variable
describing the general relativistic remnant of the gauge freedom in
clock synchronization. However now each solution of Einstein's
equations dynamically determines a preferred notion of instantaneous
3-spaces. The linearization of this canonical formulation in the
weak field approximation will allow to find Hamiltonian
Post-Minkowskian gravitational waves with an asymptotic background
and without Post-Newtonian expansion in non-harmonic 3-orthogonal
gauges.

\bigskip

Talk at the Workshop on {\it Gravitational Waves Detection with Atom
Interferometry}, held in Firenze at the Galileo Galilei Institute
for Theoretical Physics, February 23-24, 2009

\end{abstract}

\maketitle

\vfill\eject

Due to the developments in atom interferometry proposals are
appearing for using them as detectors of gravitational waves. While
this is an exciting challenge at the experimental level, it raises
theoretical problems concerning the framework to be used to put
together atomic physics, special relativity and gravity. All
experiments on the Earth are using a non-relativistic formulation of
quantum mechanics, with special and general relativistic corrections
when extended to space near Earth (see GPS, the experiment on the
Space Station, geodesy satelites), and a post-Newtonian description
of gravity seen as an external potential to be added to the
Schroedinger equation. Atoms and photons are assumed to move on
geodesics which strictly speaking do not exist in the
non-relativistic context (see for instance Ref.\cite{a1}). As a
consequence it would be desirable to have a relativistic framework
where all the actors of the game are consistently defined at least
at the classical level with a subsequent introduction of
quantization for matter (no accepted formulation of quantum gravity
exists).\medskip

Regarding atoms, usually they are described as quantum particles
with some structure in a sector of quantum field theory with a fixed
number of particles. Moreover instead of treating hem as
relativistic bound states, they are approximated with
non-relativistic quantum particles. However, non-relativistic
quantum mechanics and the theory of entanglement, with the
associated foundational problems connected to quantum non-locality,
are formulated in Galilei space-time. In it both Newtonian time and
the Euclidean 3-space, with the associated notion of spatial
distance, are {\it absolute} and Maxwell equations do not exist. The
'photons' in the discussions about atom interferometry, entanglement
and teleportation are only states with two polarizations in a
two-dimensional Hilbert space: their carrier cannot be a ray of
light in the eikonal approximation moving along a null geodesic. The
existing inclusion of electro-magnetism at the order $1/c$ made by
atomic physics destroys the Galilei group and does not allow a
consistent definition of the Poincare' one, namely a consistent
special relativistic formulation of atomic physics in Minkowski
space-time.
\bigskip

Therefore the first problem is to find a classical description of
charged particles and of the electro-magnetic field in Minkowski
space-time. Then the theory has to be reformulated in Einstein's
general relativity in asymptotically Minkowskian space-times so to
have a geometrical description of gravity with a smooth limit in the
weak field approximation to Minkowski space-time (Post-Minkowskian
gravity) and then to Galilei space-time with the Post-Newtonian
expansions. \medskip

Once this classical framework is under control, one must understand
how to include the quantum aspects of atomic physics. The first step
is to develop a consistent description of relativistic quantum
mechanics in Minkowski space-time, in which atoms and photons are
simulated by massive and massless particles with some spin structure
respectively. Moreover there must be a well defined non-relativistic
limit for massive particles. Then a phenomenological inclusion of
gravity could be done by having the special relativistic
quasi-classical (mean value of position operators in a scheme like
the one used in Eherenfest theorem) world-lines of atoms and photon
replaced with Post-Minkowskian time-like and null geodesics and with
suitable Post-Newtonian potentials describing the mutual
gravitational interactions of particles.
\bigskip

Here I review the status of the construction of this framework and
its open problems. Since we are still far away from a complete
control, it is difficult to say which could be the implications at
the experimental level at this stage.

\bigskip

The main problem in going from Galilei space-time to Minkowski one
and then to Einstein's ones (or at least at their weak field
approximation) is the absence of an intrinsic notion of {\it
instantaneous 3-space} where to give the Cauchy data at least for
Maxwell equations, so to have predictability for future times.
Already in special relativity, where the space-time is absolute,
there is no notion of simultaneity, of instantaneous 3-spaces and of
spatial distance. The light postulates say that the two-way (or
round-trip; only one clock is involved) velocity of light is a)
isotropic and b) constant (a standard constant $c$ replaces the
standard of length in existing relativistic metrology). The one-way
velocity of light between two observers depends on how their clocks
are synchronized (in general is not isotropic and point-dependent).
Usually one uses Einstein's convention for clock synchronization: an
inertial observer A send a ray of light at $x^o_i$ towards the (in
general accelerated) observer B; the ray is reflected towards A at a
point P of B world-line and then reabsorbed by A at $x^o_f$; by
convention P is synchronous with the mid-point between emission and
absorption on A's world-line, i.e. $x^o_P = x^o_i + {1\over 2}\,
(x^o_f - x^o_i)$. This convention selects the Euclidean
instantaneous 3-spaces $x^o = ct = const.$ of the inertial frames
centered on A. Only in this case the one-way velocity of light
between A and B coincides with the two-way one, $c$. However if the
observer A is accelerated the convention breaks down, because if
{\it only} the world-line of the accelerated observer A (the {\it
1+3 point of view}) is given, then the only way for defining
instantaneous 3-spaces is to identify them with the Euclidean
tangent planes orthogonal to the 4-velocity of the observer (the
local rest frames). But these planes (they are tangent spaces not
3-spaces!) will intersect each other at a distance from A's
world-line of the order of the acceleration lengths of A, so that
all the either linearly or rotationally accelerated frames, centered
on accelerated observers, based either on Fermi coordinates or on
rotating ones will develop {\it coordinate singularities}. Therefore
their instantaneous 3-spaces cannot be used for a well-posed Cauchy
problem for Maxwell equations. See Refs\cite{a2,a3,a4} for more
details and a rich bibliography on these topics.
\bigskip

In general relativity, where the space-time becomes dynamical and
the equivalence principle forbids the existence of global inertial
frames, the problem of {\it clock synchronization} becomes
unavoidable and fundamental. Already near the Earth there is the
ACES mission of ESA \cite{a5}, programmed for 2013, which will make
possible a measurement of the gravitational redshift of the Earth
from the two-way link among a microwave clock (PHARAO) on the Space
Station and similar clocks on the ground: the proposed microwave
link should make possible the control of effects on the scale of 5
picoseconds. This will be a test of post-Newtonian gravity in the
framework of Einstein's geometrical view of gravitation: the
redshift is a measure of the $1/c^2$ deviation of post-Newtonian
null geodesics from Minkowski ones.
\bigskip

As a consequence we need a fully relativistic formulation of the
classical background of atomic physics, considered as an effective
theory of positive-energy scalar (or spinning) particles with mutual
Coulomb interaction plus the transverse electro-magnetic field in
the radiation gauge, valid for energies below the threshold of pair
production. To build the Poincare' generators of such an isolated
system we have to face the problem of clock synchronization
(determination of an instantaneous 3-space), the problem that the
Poincare' boosts, differently from the Galilei ones, are
interaction-dependent and the old problem of the relativistic
extension of the Newtonian center of mass. Moreover we need a
formulation in which the change of the clock synchronization
convention does not alter the physical results, namely it must be
formulated as a gauge transformation of the theory \cite{a3}.
Clearly, this theory must put inertial and non-inertial frames of
Minkowski space-time on the same level: this will allow the
transition to general relativity, where global inertial frames are
absent and the instantaneous 3-spaces are  dynamical \cite{a6},
since each solution of Einstein's equations dynamically selects a
preferred clock synchronization convention.
\bigskip

{\it Parametrized Minkowski theories} \cite{a7}, \cite{a3},
\cite{a2}, with the associated {\it inertial and non-inertial
rest-frame instant form of dynamics}, solve these problems and,
together with the results of Refs.\cite{a8,a9,a10}, allow to get a
formulation of relativistic atomic physics \cite{a11}, \cite{a12},
\cite{a4}, both in inertial and non-inertial frames of Minkowski
space-time. To formulate this theory without the coordinate
singularities of the 1+3 point of view, we need the {\it 3+1 point
of view} \cite{a3}, in which we assign: a) the world-line of an
arbitrary time-like observer; b) an admissible 3+1 splitting of
Minkowski space-time, namely a nice foliation with space-like
instantaneous 3-spaces (i.e. a clock synchronization
convention).\medskip

This allows to define a {\it global non-inertial frame} centered on
the observer and to use observer-dependent Lorentz-scalar {\it radar
4-coordinates} $\sigma^A = (\tau ;\sigma^r)$, where $\tau$ is a
monotonically increasing function of the proper time of the observer
and $\sigma^r$ are curvilinear 3-coordinates on the 3-space
$\Sigma_{\tau}$ having the observer as origin. If $x^{\mu} \mapsto
\sigma^A(x)$ is the coordinate transformation from the inertial
Cartesian 4-coordinates $x^{\mu}$ to radar coordinates, its inverse
$\sigma^A \mapsto x^{\mu} = z^{\mu}(\tau ,\sigma^r)$ defines the
embedding functions $z^{\mu}(\tau ,\sigma^r)$ describing the
3-spaces $\Sigma_{\tau}$ as embedded 3-manifold into Minkowski
space-time. The induced 4-metric on $\Sigma_{\tau}$ is the following
functional of the embedding $g_{AB}(\tau ,\sigma^r) = [z^{\mu}_A\,
\eta_{\mu\nu}\, z^{\nu}_B](\tau ,\sigma^r)$, where $z^{\mu}_A =
\partial\, z^{\mu}/\partial\, \sigma^A$ and $\eta_{\mu\nu} = \sgn\,
(+---)$ is the flat metric ($\sgn = \pm 1$ according to either the
particle physics $\sgn = 1$ or the general relativity $\sgn = - 1$
convention). While the 4-vectors $z^{\mu}_r(\tau ,\sigma^u)$ are
tangent to $\Sigma_{\tau}$, so that the unit normal $l^{\mu}(\tau
,\sigma^u)$ is proportional to $\epsilon^{\mu}{}_{\alpha
\beta\gamma}\, [z^{\alpha}_1\, z^{\beta}_2\, z^{\gamma}_3](\tau
,\sigma^u)$, we have $z^{\mu}_{\tau}(\tau ,\sigma^r) = [N\, l^{\mu}
+ N^r\, z^{\mu}_r](\tau ,\sigma^r)$ ($N(\tau ,\sigma^r) = \sgn\,
[z^{\mu}_{\tau}\, l_{\mu}](\tau ,\sigma^r)$ and $N_r(\tau ,\sigma^r)
= - \sgn\, g_{\tau r}(\tau ,\sigma^r)$ are the lapse and shift
functions).\medskip

The foliation is nice and admissible if it satisfies the conditions:
\hfill\break
 1) $N(\tau ,\sigma^r) > 0$ in every point of
$\Sigma_{\tau}$ (the 3-spaces never intersect);\hfill\break
 2) $\sgn\, g_{\tau\tau}(\tau ,\sigma^r) > 0$, so to avoid the
 coordinate singularity of the rotating disk, and with the positive-definite 3-metric
$h_{rs}(\tau ,\sigma^u) = - \sgn\, g_{rs}(\tau ,\sigma^u)$ having
three positive eigenvalues (these are the M$\o$ller conditions
\cite{a2,a4});\hfill\break
 3) all the 3-spaces $\Sigma_{\tau}$ must tend to the same space-like
 hyper-plane at spatial infinity (so that there are always asymptotic inertial
observers to be identified with the fixed stars).\hfill\break
 These conditions imply that global {\it rigid} rotations are
forbidden in relativistic theories \cite{a2}.\medskip

The 4-metric $g_{AB}(\tau ,\vec \sigma )$ on $\Sigma_{\tau}$ has the
components $\sgn\, g_{\tau\tau} = N^2 - N_r\, N^r$, $- \sgn\,
g_{\tau r} = N_r = h_{rs}\, N^s$, $h_{rs} = - \sgn\, g_{rs} =
\sum_{a=1}^3\, e_{(a)r}\, e_{(a)s} = \gamma^{1/3}\, \sum_{a=1}^3\,
e^{2\, \sum_{\bar b =1}^2\, \gamma_{\bar ba}\, R_{\bar b}}\,
V_{ra}(\theta^i)\, V_{sa}(\theta^i)$), where $e_{(a)r}(\tau
,\sigma^u)$ are cotriads on $\Sigma_{\tau}$, $\gamma (\tau
,\sigma^r) = det\, h_{rs}(\tau ,\sigma^r)$ is the 3-volume element
on $\Sigma_{\tau}$, $\lambda_a(\tau ,\sigma^r) = [\gamma^{1/6}\,
e^{\sum_{\bar b =1}^2\, \gamma_{\bar ba}\, R_{\bar b}}](\tau
,\sigma^r)$ are the positive eigenvalues of the 3-metric
($\gamma_{\bar aa}$ are suitable numerical constants) and
$V(\theta^i(\tau ,\sigma^r))$ are diagonalizing rotation matrices
depending on three Euler angles. The components $g_{AB}$ or the
quantities $N$, $N_r$, $\gamma$, $R_{\bar a}$, $\theta^i$, play the
role of the {\it inertial potentials} generating the relativistic
apparent forces in the non-inertial frame. It can be shown \cite{a4}
that the Newtonian inertial potentials are hidden in the functions
$N$, $N_r$ and $\theta^i$.

\bigskip

In parametrized Minkowski theories one considers any isolated system
(particles, strings, fields, fluids) admitting a Lagrangian
description, because it allows, through the coupling to an external
gravitational field, the determination of the matter energy-momentum
tensor and of the ten conserved Poincare' generators $P^{\mu}$ and
$J^{\mu\nu}$ (assumed finite) of every configuration of the system.
Then one replaces the external gravitational 4-metric in the coupled
Lagrangian with the 4-metric $g_{AB}(\tau ,\sigma^r)$ of an
admissible 3+1 splitting of Minkowski space-time and  the matter
fields with new ones knowing the instantaneous 3-spaces
$\Sigma_{\tau}$. For instance a Klein-Gordon field $\tilde \phi (x)$
will be replaced with $\phi(\tau ,\sigma^r) = \tilde \phi (z(\tau
,\sigma^r))$; the same for every other field. Instead for a
relativistic particle with world-line $x^{\mu}(\tau )$ we must make
a choice of its energy sign: then it will be described by
3-coordinates $\eta^r(\tau )$ defined by the intersection of the
world-line with $\Sigma_{\tau}$: $x^{\mu}(\tau ) = z^{\mu}(\tau
,\eta^r(\tau ))$. In this way we get a Lagrangian depending on the
given matter and on the embedding $z^{\mu}(\tau ,\sigma^r)$, which
is invariant under frame-preserving diffeomorphisms. As a
consequence, there are four first-class constraints (an analogue of
the super-Hamiltonian and super-momentum constraints of canonical
gravity) implying that the embeddings $z^{\mu}(\tau ,\sigma^r)$ are
{\it gauge variables}, so that all the admissible non-inertial
frames are gauge equivalent, namely physics does {\it not} depend on
the clock synchronization convention and on the choice of the
3-coordinates $\sigma^r$: only the appearances of phenomena change
by changing the notion of instantaneous 3-space.

\bigskip

The {\it inertial rest-frame instant form of dynamics for isolated
systems} is obtained by choosing the 3+1 splitting corresponding to
the intrinsic inertial rest frame of the isolated system centered on
an inertial observer: the instantaneous 3-spaces, named Wigner
3-space due to the fact that the 3-vectors inside them are Wigner
spin-1 3-vectors \cite{a3,a7}, are orthogonal to the conserved
4-momentum $P^{\mu}$ of the configuration. In Ref.\cite{a4} there is
the extension to admissible non-inertial rest frames, where
$P^{\mu}$ is orthogonal to the asymptotic space-like hyper-planes to
which the instantaneous 3-spaces tend at spatial infinity. This
non-inertial family of 3+1 splittings is the only one admitted by
the asymptotically Minkowskian space-times covered by canonical
gravity formulation discussed at the end of the paper.
\bigskip

In the inertial rest frames there are {\it only} three notions of
collective variables, which can be built by using {\it only} the
Poincare' generators (they are {\it non-local} quantities knowing
the whole $\Sigma_{\tau}$) \cite{a8}: the canonical non-covariant
Newton-Wigner center of mass (or center of spin), the non-canonical
covariant Fokker-Pryce center of inertia and the non-canonical
non-covariant M$\o$ller center of energy. All of them tend to the
Newtonian center of mass in the non-relativistic limit. See
Ref.\cite{a3} for the M$\o$ller non-covariance world-tube around the
Fokker-Pryce 4-vector identified by these collective variables. As
shown in Refs.\cite{a8,a9,a10} these three variables can be
expressed as known functions of the rest time $\tau$, of the
canonically conjugate Jacobi data (frozen Cauchy data) $\vec z =
Mc\, {\vec x}_{NW}(0)$ (${\vec x}_{NW}(\tau )$ is the standard
Newton-Wigner 3-position) and $\vec h = \vec P/Mc$, of the invariant
mass $Mc = \sqrt{\sgn\, P^2}$ of the system and of its rest spin
${\vec {\bar S}}$. It is convenient to center the inertial rest
frame on the Fokker-Pryce inertial observer. As a consequence, every
isolated system (i.e. a closed universe) can be visualized as a
decoupled non-covariant collective (non-local) pseudo-particle
described by the frozen Jacobi data $\vec z$, $\vec h$ carrying a
{\it pole-dipole structure}, namely the invariant mass and the rest
spin of the system, and with an associated external realization of
the Poincare' group. The universal breaking of Lorentz covariance is
connected to this decoupled non-local collective variable and is
irrelevant because all the dynamics of the isolated system leaves
inside the Wigner 3-spaces and is Wigner-covariant. In each Wigner
3-space $\Sigma_{\tau}$ there is a unfaithful internal realization
of the Poincare' algebra, whose generators are built by using the
energy-momentum tensor of the isolated system. While the internal
energy and angular momentum are $Mc$ and ${\vec {\bar S}}$
respectively, the internal 3-momentum vanishes: it is the rest frame
condition. Also the internal Lorentz boost (whose expression in
presence of interactions is given for the first time) vanishes: this
condition identifies the covariant non-canonical Fokker-Pryce center
of inertia as the natural inertial observer origin of the
3-coordinates $\sigma^r$ in each Wigner 3-space. As a consequence
\cite{a11} there are three pairs of second class
(interaction-dependent) constraints eliminating the internal
3-center of mass and its conjugate momentum inside the Wigner
3-spaces \cite{a12}: this avoids a double counting of the collective
variables and allows to re-express the dynamics only in terms of
internal Wigner-covariant relative variables. In the case of
relativistic particles the reconstruction of their world-lines
requires a complex interaction-dependent procedure delineated in
Ref.\cite{a10}. See Ref.\cite{a11} for the comparison with the other
formulations of relativistic mechanics developed for the study of
the problem of {\it relativistic bound states}.

\bigskip

In Refs.\cite{a9} and \cite{a11}, there is the formulation of
relativistic atomic physics in the inertial rest frames (extended to
the non-inertial ones in Ref.\cite{a4}). This was possible because
we considered Grassmann-valued electric charges for the particles
($Q_i^2 = 0$, $Q_i\, Q_j = Q_j\, Q_i \not= 0$ for $i \not= j$),
which give rise to a two-level charge structure after quantization.
This allows\hfill\break
 a) to make an ultraviolet regularization of Coulomb
self-energies;\hfill\break
 b) to make an infrared regularization
eliminating the photon emission;\hfill\break
 c) to express the Lienard-Wiechert potentials only in terms of the 3-coordinates
$\eta^r_i(\tau )$ and the conjugate 3-momenta $\kappa_{ir}(\tau)$ in
a way independent from the used (retarded, advanced,..) Green
function.\medskip

All this amounts to reformulate the dynamics of the one-photon
exchange Feynman diagram as a Cauchy problem with well defined
classical potentials. Moreover there is a canonical transformation
\cite{a11} sending the above system in a transverse radiation field
(in- or out-fields) decoupled, in the global rest frame, from
Coulomb-dressed particles with a mutual interaction described by the
sum of the Coulomb potential plus the Darwin potential. Therefore
for the first time we are able to obtain results, previously derived
from instantaneous approximations to the Bethe-Salpeter equation for
the description of relativistic bound states (see the bibliography
of Ref.\cite{a9}), starting from the classical theory. Moreover, for
the first time, at least at the classical level, we have been able
to avoid the Haag theorem according to which the interaction picture
does not exist in QFT.\bigskip

Therefore now we have an acceptable special relativistic description
of massive particles and of the electro-magnetic field in the
radiation gauge. We are now developing the rest-frame description of
massless particles with helicity \cite{a13} (see also Appendix C of
Ref.\cite{a14}) to simulate "photons" as an eikonal geometrical
optic approximation of rays of light. Also a quasi-classical
description of a relativistic two-level atom is under study
\cite{a15}. These ill be the building blocks needed to try to give a
relativistic description of the classical background of an atom
interferometer.

\bigskip

Before introducing gravity let us consider the problem of quantizing
the inertial and non-inertial rest frame instant form of dynamics in
special relativity. In refs.\cite{a16} there is the quantization of
positive-energy free scalar and spinning particles in a family of
non-inertial rest frames where the instantaneous 3-spaces are
space-like hyper-planes with differentially rotating coordinates. We
take the point of view {\it not to quantize the inertial effects}
(the appearances of phenomena): the embedding $z^{\mu}(\tau
,\sigma^r)$ remains a c-number and we get results compatible with
atomic spectra. Instead the quantization of {\it fields} in
non-inertial frames is an open problem due to the no-go theorem of
Ref.\cite{a17} showing the existence of obstructions to the unitary
evolution of a massive quantum Klein-Gordon field between two
space-like surfaces of Minkowski space-time. Its solution, i.e. the
identification of all the 3+1 splittings allowing unitary evolution,
will be a prerequisite to any attempt to quantize canonical gravity
taking into account the equivalence principle (global inertial
frames do not exist!).

\bigskip

Let us make a digression on relativistic entanglement.  The
formulation of entanglement in non-relativistic quantum mechanics is
based on the assumption (the zeroth postulate) that a composite
system with two (or more) subsystems is described by a Hilbert space
which is the {\it tensor product} of the Hilbert spaces of the
subsystems: $H = H_1 \otimes H_2$. When interactions between the
subsystems are present, one makes a unitary transformation to $H =
H_1 \otimes H_2 = H_{com} \otimes H_{rel}$, where the Hilbert spaces
$H_{com}$ and $H_{rel}$ describe the decoupled Newtonian center of
mass of the two subsystems and their relative motion respectively.
Then $H_{com}$ can be replaced with a frozen Hamilton-Jacobi Hilbert
space $H_{com, HJ}$, containing only non-evolving Jacobi data for
the center of mass, by a unitary transformation.\medskip

Due to the non-locality of the canonical non-covariant
(Newton-Wigner) center of mass, described by the frozen Jacobi data
$\vec z$ and $\vec h$, the previous type of unitary equivalence
breaks down in relativistic quantum mechanics and therefore in
relativistic quantum atomic physics. As shown in Ref.\cite{a14}, the
only consistent relativistic quantization is based on the Hilbert
space $H = H_{com, HJ} \otimes H_{rel}$, which has the correct
non-relativistic limit. The frozen nature of the center of mass
avoids the violation of relativistic causality implied by the
Hegerfeldt theorem \cite{a18} about the instantaneous spreading of
the wave packets.\medskip

The breaking of the zeroth postulate at the relativistic level, $H
\not= H_1 \otimes H_2$, is due to the fact that, already in the
non-interacting case, in the tensor product of two quantum
Klein-Gordon fields, $\phi_1(x_1)$ and $\phi_2(x_2)$,  most of the
states correspond to configurations in Minkowski space-time in which
one particle may be present in the absolute future of the other
particle. This is due to the fact that the two times $x^o_1$ and
$x^o_2$ are totally uncorrelated, or in other words there is {\it no
notion of instantaneous 3-space} (clock synchronization convention).
Also the scalar products in the two formulations are completely
different as shown in Ref.\cite{a19}. In S-matrix theory this
problem is eliminated by avoiding the interpolating states at finite
(the problem of the Haag theorem) and going the the asymptotic (in
the times $x^o_i$) limit of the free in- and out- states. However in
atomic physics we need interpolating states, and not S-matrix, to
describe a laser beam resonating in a cavity and intersected by a
beam of atoms!\medskip

The non validity of the zeroth postulate for composite systems
implies that Einstein's notion of separability is not valid since in
$H = H_{com, HJ} \otimes H_{rel}$ the composite system must be
described by means of relative variables in a Wigner 3-space (this
is a type of weak form of  relationism different from the
formulations connected to the Mach principle). As a consequence
every component of an isolated relativistic quantum system is
entangled with every other component, even if it is not causally
connected. These {\it kinematical non-locality} and  {\it
kinematical spatial non-separability} introduced by special
relativity reduce the relevance of {\it quantum non-locality} in the
study of the foundational problems of quantum mechanics \cite{a20}
which have to be rephrased in terms of relative variables. Moreover,
the control of Poincare' kinematics will force to reformulate the
experiments connected with Bell inequalities and teleportation in
terms of isolated systems containing: a) the observers with their
measuring apparatus (Alice and Bob as macroscopic quasi-classical
objects); b) the particles of the protocol (but now the ray of
light, the "photons" carrying the polarization, move along null
geodesics); c) the environment (macroscopic either quantum or
quasi-classical object).

\bigskip

Finally let us consider the reformulation of the previous results in
the canonical formulation of metric and tetrad gravity given in
Refs.\cite{12,a}, which is the final version of previous works in
Refs.\cite{13}. In Einstein's general relativity space-time is
dynamical, gravity is described by the 4-metric tensor and the
equivalence principle says that global inertial frames do not exist.
To get its Hamiltonian formulation we must use the same 3+1
formalism previously introduced for parametrized Minkowski theories
and replace the Einstein-Hilbert action with the ADM one. However
now the basic dynamical variable is the 4-metric $g_{AB}(\tau,
\sigma^r)$ and not the embedding $z^{\mu}(\tau, \sigma^r)$
($z^{\mu}_A$ are now the transformation coefficients for tensors
from  world 4-coordinates $x^{\mu}$ to adapted radar 4-coordinates).
Since tetrad gravity is more natural for the coupling to the
fermions, the 4-metric is decomposed in terms of cotetrads, $g_{AB}
= E_A^{(\alpha)}\, \eta_{(\alpha)(\beta)}\, E^{(\beta)}_B$
($(\alpha)$ are flat indices), and the ADM action, now a functional
of the 16 fields $E^{(\alpha)}_A(\tau, \sigma^r)$, is taken as the
action for ADM tetrad gravity. This leads to an interpretation of
gravity based on a congruence of time-like observers endowed with
orthonormal tetrads: in each point of space-time the time-like axis
is the  unit 4-velocity of the observer, while the spatial axes are
a (gauge) convention for observer's gyroscopes.
\medskip

This canonical formulation holds in a special class of globally
hyperbolic, asymptotically Minkowskian at spatial infinity,
topologically trivial space-times with boundary conditions killing
super-translations so that the asymptotic symmetries are reduced to
the ADM Poincare' group as shown in Refs.\cite{13}. The ADM energy
turns out to be the Hamiltonian of ADM tetrad gravity.\medskip

The absence of super-translations implies that the non-inertial rest
frames are the only family of 3+1 splittings admitted by these
asymptotically Minkowskian space-times. Therefore the instantaneous
3-spaces are asymptotically orthogonal to the ADM 4-momentum, are
non-inertial rest frames of the 3-universe and admit asymptotic
inertial observers (the fixed stars). As a consequence the
3-universe contained in an instantaneous 3-space can be described as
a decoupled non-covariant non-observable pseudo-particle carrying a
pole-dipole structure, whose mass and spin are those identifying the
configuration of the "gravitational field plus matter" isolated
system present in the 3-universe.\medskip

Moreover, if we switch down the Newton constant, we get the
description of the matter present in these space-times in the
non-inertial rest frames of Minkowski space-time (deparametrization
of general relativity) with the ADM Poincare' group collapsing in
the Poincare' group of particle physics.

\medskip

The kinematical Poincare' group connecting inertial frames in
special relativity and its enlargement to the group of
frame-preserving diffeomorphisms required for the treatment of
non-inertial frames in parametrized Minkowski theories are now
replaced by the full spatio-temporal diffeonorphism group enlarged
with the O(3,1) gauge group of the Newman-Penrose approach (the
extra gauge freedom acting on the tetrads in the tangent space of
each point of space-time and reducing from 16 to 10 the number of
independent fields like in metric gravity). The relativity principle
of special relativity is replaced with the principle of general
covariance (invariance in form of physical laws).

\medskip

Since in general relativity the whole chrono-geometrical structure
of space-time, described by the 4-metric and the associated line
element, is {\it dynamical}, it turns out that every solution of
Einstein equations {\it dynamically} selects its preferred
instantaneous 3-spaces (modulo coordinate transformations)
\cite{a6}. As a consequence also the clock synchronization
convention acquires a dynamical character. The gravitational field,
i.e. the 4-metric, is not only the potential of the gravitational
interaction but it also teaches relativistic causality to the other
fields (it says to each massless particle which are the allowed
trajectories in each point). This geometrical property is lost when
the 4-metric is split in a background plus a perturbation (like in
quantum field theory and string theory for being able to define a
Fock space), since the chrono -geometrical structure is frozen to
the one of the background. In Refs.\cite{12,a} such a splitting is
never done, since there is an {\it asymptotic} Minkowskian
background.

\bigskip

Let us remark that in the ADM Hamiltonian formulation of general
relativity in the York canonical basis of Ref.\cite{12} we use the
same decomposition of the 4-metric $g_{AB}$ in terms of the
quantities $N$, $N_r$, $\gamma$, $R_{\bar a}$, $\theta^i$, like in
the non-inertial frames of special relativity. However now we have
that:\medskip

a) the quantities $R_{\bar a}(\tau ,\sigma^r)$, $\bar a =1,2$,
become the physical {\it tidal} degrees of freedom of the
gravitational field (the polarizations of the gravitational waves in
the linearized theory);

b) the lapse and shift functions $N$ and $N_r$ and the angles
$\theta^i$ (determining the 3-coordinates $\sigma^r$ on
$\Sigma_{\tau}$) remain inertial gauge variables, namely they play
the role of inertial potentials like in special relativity;

c) the 3-volume element $\gamma (\tau ,\sigma^r)$ is determined by
the super-Hamiltonian constraint (the Lichnerowicz equation) in
terms of the other variables;

d) there is an extra inertial potential (replacing $\gamma$ of
special relativity) describing the non-dynamical part of the freedom
in choosing the clock synchronization convention (once this gauge
freedom is fixed the final shape of the instantaneous 3-space is
dynamically determined; it is the remnant of the special
relativistic gauge freedom), i.e. the trace $K(\tau ,\sigma^r)$ of
the extrinsic curvature (also named the York time) of the
non-Euclidean 3-space $\Sigma_{\tau}$, which is a functional of
$g_{AB}$ in special relativity and has {\it no Newtonian
counterpart}; in the York canonical basis it is the only gauge
variable described by a momentum (a reflex of the Lorentz signature
of space-time) and it gives rise to a negative kinetic term in the
weak ADM energy vanishing only in the gauges ${}^3K(\tau, \vec
\sigma) = 0$;

e) finally the extra O(3,1) gauge freedom for the tetrads (the gauge
freedom for each observer to choose three gyroscopes as spatial axes
and to choose the law for their transport along the world-line) can
be fixed by choosing tetrads adapted to the 3+1 splitting (the
time-like tetrad is the unit normal to the 3-space) with the
so-called Schwinger time gauges.

\bigskip

In Ref.\cite{a} we study the coupling of N charged scalar particles
plus the electro-magnetic field to ADM tetrad gravity  in this class
of asymptotically Minkowskian space-times without
super-translations. To regularize the self-energies both the
electric charge and the sign of the energy of the particles are
Grassmann-valued. The introduction of the non-covariant radiation
gauge allows to reformulate the theory in terms of transverse
electro-magnetic fields and to extract the generalization of the
Coulomb interaction among the particles in the Riemannian
instantaneous 3-spaces of global non-inertial frames.

\medskip

After the reformulation of the whole system in the York canonical
basis, we give the restriction of the Hamilton equations and of the
constraints to the family of {\it non-harmonic 3-orthogonal}
Schwinger time gauges, in which the instantaneous Riemannian
3-spaces have a non-fixed trace ${}^3K$ of the extrinsic curvature
but a diagonal 3-metric.
\medskip

\bigskip

Starting from the results obtained in the family of non-harmonic
3-orthogonal Schwinger gauges, it  will be possible to define a
linearization of ADM canonical tetrad gravity plus matter in the
weak field approximation, to obtain a formulation of Hamiltonian
Post-Minkowskian gravity (without Post-Newtonian expansions) with
non-flat Riemannian 3-spaces and asymptotic Minkowski background:
i.e. with a decomposition of the 4-metric tending to the asymptotic
Minkowski metric at spatial infinity, $g_{AB} = \eta_{AB} + h_{AB}\,
\rightarrow \eta_{AB}$ (the small perturbation $h_{AB}$ vanishes at
spatial infinity). We will show that a consequence of this approach
is the possibility of describing part (or maybe all) dark matter as
a {\it relativistic inertial effect} determined by the gauge
variable ${}^3K(\tau, \sigma^r)$ (not existing in Newtonian gravity,
where the Euclidean 3-space is an absolute notion): the rotation
curves of galaxies would then experimentally determine a preferred
choice of the instantaneous 3-spaces.
\medskip

It is at his stage that it will be possible to see how to try to
simulate the classical background of atom interferometry in presence
of Post-Newtonian gravity as it is done in Ref.\cite{a1}.

\bigskip

Finally, if we will replace the matter with a perfect fluid (for
instance dust), this  will alow us to try to see whether the York
canonical basis can help in developing the back-reaction \cite{b}
approach to dark energy, according to which dark energy is a
byproduct of the non-linearities of general relativity when one
considers mean values on large scales.

\vfill\eject


\begin{thebibliography}{}

\bibitem{a1}S.Dimopoulous, P.W.Graham, J.M.Hogan and M.A.Kasevich, {\it General
 Relativistic Effects in Atom Interferometry} (arXiv:
 0802.4098).\hfill\break
 S.Dimopoulous, P.W.Graham, J.M.Hogan, M.A.Kasevich and S.Rajendran,
 {\it An Atomic Gravitational Wave Interferometric Sensor (AGIA)}
 (arXiv: 0806.2125).


\bibitem{a2}D.Alba and L.Lusanna, {\it Generalized Radar 4-Coordinates
and Equal-Time Cauchy Surfaces for Arbitrary Accelerated Observers}
(2005),  Int.J.Mod.Phys. {\bf D16}, 1149 (2007) (gr-qc/0501090).

\bibitem{a3}L.Lusanna, {\it The Chrono-Geometrical Structure of Special and General
Relativity: A Re-Visitation of Canonical Geometrodynamics}, lectures
at 42nd Karpacz Winter School of Theoretical Physics: Current
Mathematical Topics in Gravitation and Cosmology, Ladek, Poland,
6-11 Feb 2006, Int.J.Geom.Methods in Mod.Phys. {\bf 4}, 79 (2007).
(gr-qc/0604120).\hfill\break
 L.Lusanna, {\it The Chronogeometrical
Structure of Special and General Relativity: towards a
Background-Independent Description of the Gravitational Field and
Elementary Particles} (2004), in {\it General Relativity Research
Trends}, ed. A.Reiner, Horizon in World Physics vol. 249 (Nova
Science, New York, 2005) (gr-qc/0404122).


\bibitem{a4}D.Alba and L.Lusanna, {\it Charged Particles and the
Electro-Magnetic Field in Non-Inertial Frames}, to appear in
Int.J.Geom. Methods in Modern Phys. (arXiv 0812.3057).



\bibitem{a5} L.Cacciapuoti and C.Salomon, {\it ACES: Mission Concept and
Scientific Objective}, 28/03/2007, ESA document, Estec
(ACES{}\_{}Science{}\_{}v1{}\_{}printout.doc).\hfill\break
 L.Blanchet, C.Salomon, P.Teyssandier and P.Wolf, {\it Relativistic
Theory for Time and Frequency Transfer to Order $1/c^3$},
Astron.Astrophys. {\bf 370}, 320 (2000).\hfill\break
 L. Duchayne, F. Mercier and P. Wolf, {\it Orbitography for Next
Generation Space Clocks}, 2007, (arXiv:0708.2387).\hfill\break
 L.Lusanna, {\it Dynamical Emergence of 3-Space in General
Relativity: Implications for the ACES Mission}, in Proc. of the 42th
Rencontres de Moriond {\it Gravitational Waves and Experimental
Gravity}, La Thuile (Italy), 11-18 March 2007.\hfill\break
 See also  the talks at the {\it SIGRAV Graduate School on Experimental
Gravitation in Space}(Firenze, September 25-27, 2006)
(http://www.fi.infn.it/GGI-grav-space/egs{}\_{}s.html); at the
Workshop {\it Advances in Precision Tests and Experimental
Gravitation in Space} (Firenze, September 28/30, 2006)
(http://www.fi.infn.it/GGI-grav-space/egs{}\_{}w.html); at the
Workshop "Theoretical Aspects of the ACES Mission" (Firenze, April
29-30, 2008) (ftp://cacciapuoti:In73rn0@ftp.estec.esa.int/ ); at the
Workshop on "ACES and Future GNSS-based Earth Observation and
Navigation" (Muenchen, May 26-27, 2008)
(http://www.iapg.bv.tum.de/12735--~aces~programme.html).


\bibitem{a6} L.Lusanna and M.Pauri,
{\it Explaining Leibniz equivalence as difference of non-inertial
Appearances: Dis-solution of the Hole Argument and physical
individuation of point-events}, History and Philosophy of Modern
Physics {\bf 37}, 692 (2006) (gr-qc/0604087); {\it The Physical Role
of Gravitational and Gauge Degrees of Freedom in General Relativity.
I: Dynamical Synchronization and Generalized Inertial Effects; II:
Dirac versus Bergmann Observables and the Objectivity of
Space-Time}, Gen.Rel.Grav. {\bf 38}, 187 and 229 (2006)
(gr-qc/0403081 and 0407007); {\it Dynamical Emergence of
Instantaneous 3-Spaces in a Class of Models of General Relativity},
to appear in the book {\it Relativity and the Dimensionality of the
World}, ed. A. van der Merwe (Springer Series Fundamental Theories
of Physics) (gr-qc/0611045).



\bibitem{a7}L. Lusanna, {\it The N- and 1-Time Classical Descriptions of N-Body
Relativistic Kinematics and the Electromagnetic Interaction}, Int.
J. Mod. Phys. {\bf A12}, 645 (1997).

\bibitem{a8} D.Alba, L.Lusanna and M.Pauri, \textit{New Directions in
Non-Relativistic and Relativistic Rotational and Multipole
Kinematics for N-Body and Continuous Systems} (2005), in
\textit{Atomic and Molecular Clusters: New Research}, ed.Y.L.Ping
(Nova Science, New York, 2006) (hep-th/0505005).\hfill\break
  D.Alba, L.Lusanna and M.Pauri, \textit{Centers of Mass and Rotational
Kinematics for the Relativistic N-Body Problem in the Rest-Frame
Instant Form}, J.Math.Phys. \textbf{43}, 1677-1727 (2002)
(hep-th/0102087).\hfill\break
  D.Alba, L.Lusanna and M.Pauri,
\textit{ Multipolar Expansions for Closed and Open Systems of
Relativistic Particles} , J. Math.Phys. \textbf{46}, 062505, 1-36
(2004) (hep-th/0402181).


\bibitem{a9} H.Crater and L.Lusanna, {\it The Rest-Frame Darwin Potential from
 the Lienard-Wiechert Solution in the Radiation Gauge},
 Ann.Phys.(N.Y.) {\bf 289}, 87 (2001)(hep-th/0001046).\hfill\break
 D.Alba, H.Crater and L.Lusanna, {\it The Semiclassical
 Relativistic Darwin Potential for Spinning Particles in the
 Rest-Frame Instant Form: Two-Body Bound States with Spin 1/2
 Constituents}, Int.J.Mod.Phys. {\bf A16}, 3365 (2001) (hep-th/0103109).


\bibitem{a10} D.Alba, H.W.Crater and L.Lusanna, \textit{Hamiltonian
Relativistic Two-Body Problem: Center of Mass and Orbit
Reconstruction}, J.Phys. {\bf A40}, 9585 (2007) (gr-qc/0610200).


\bibitem{a11}D.Alba, H.W.Crater and L.Lusanna, {\it Towards Relativistic
Atom Physics. I. The Rest-Frame Instant Form of Dynamics and a
Canonical Transformation for a system of Charged Particles plus the
Electro-Magnetic Field}, to appear in Canad.J.Phys. (arXiv:
0806.2383).



\bibitem{a12}D.Alba, H.W.Crater and L.Lusanna, {\it Towards Relativistic
Atom Physics. II. Collective and Relative  Relativistic Variables
for a System of Charged Particles plus the Electro-Magnetic Field},
to appear in Canad.J.Phys.(0811.0715).

\bibitem{a13} D.Alba, H.W.Crater and L.Lusanna, \textit{Massless Particles
plus Matter in the Rest-Frame Instant Form of Dynamics}, in
preparation.

\bibitem{a14}D.Alba, H.W.Crater and L.Lusanna, {\it Relativistic
Quantum Mechanics and Relativistic Entanglement in the Rest-Frame
Instant Form of Dynamics}, (arXiv 0907.1816).

\bibitem{a15} D.Alba, H.W.Crater and L.Lusanna,
A Relativistic Version of the Two-Level Atom in the Rest-Frame
Instant Form of Dynamics, in preparation.



\bibitem{a16}D.Alba and L.Lusanna, {\it Quantum Mechanics in
Non-Inertial Frames with a Multi-Temporal Quantization Scheme: I)
Relativistic Particles}, Int.J.Mod.Phys. {\bf A21}, 2781 (2006)
(hep-th/0502060)\hfill\break
 D.Alba, {\it Quantum Mechanics in Non-Inertial Frames with a
Multi-Temporal Quantization Scheme: II) Non-Relativistic Particles},
Int.J.Mod.Phys. {\bf A21}, 3917 (2006) (hep-th/0504060).




\bibitem{a17}Torre, C.G. and Varadarajan, M. {\it Functional Evolution of
Free Quantum Fields}, Clas. Quantum Grav. {\bf 16}, 2651-2668
(1999).\hfill\break
 A.D.Helfer, {\it The Hamiltonian of Linear Quantum Fields}
 (hep-th/990811).\hfill\break
 A.Arageorgis, J.Earman and L.Ruetsche, {\it Weyling the Time
 Away: the Non-Unitary Implementability of Quantum Field Dynamics
 o n Curved Spacetimes}, Studies in History and
 Philosophy of Modern Physics {\bf 33}, 151 (2002).






\bibitem{a18}G.C.Hegerfeldt, {\it Remark on Causality and Particle
Localization}, Phys.Rev. {\bf D10}, 3320 (1974). {\it Violation of
Causality in Relativistic Quantum Theory?}, Phys.Rev.Lett. {\bf 54},
2395 (1985); {\it Instantaneous Spreading and Einstein Causality in
Quantum Theory}, Ann.Phys.Lpz. {\bf 7}, 716 (1998)
(quant-ph/9809030); {\it Causality, Particle Localization and
Positivity of the Energy},  in {\it Irreversibility and Causality in
Quantum Theory - Semigroups and Rigged Hilbert Spaces}, eds. A.Bohm,
H.D.Doebner and P.Kielanowski, Lecture Notes in Physics 504, p.238
(Springer , NewYork, 1998) (quant-ph/9806036).

\bibitem{a19} G.Longhi and L.Lusanna, {\it
Bound-State Solutions, Invariant Scalar Products and Conserved
Currents for a Class of Two-Body Relativistic Systems}, Phys.Rev.
\textbf{D34}, 3707 (1986).


\bibitem{a20}M.Schlosshauer, {\it Decoherence and the Quantum-to-Classical
Transition} (Springer, Berlin, 2007); {\it Decoherence, the
Measurement Problem and Interpretations of Quantum Mechanics},
Rev.Mod.Phys. {\bf 76}, 1267 (2004).


\bibitem{12}D.Alba and L.Lusanna, {\it The York Map as a Shanmugadhasan
Canonical Transformationn in Tetrad Gravity and the Role of
Non-Inertial Frames in the Geometrical View of the Gravitational
Field}, Gen.Rel.Grav. {\bf 39}, 2149 (2007) (gr-qc/0604086, v2).

\bibitem{a}D.Alba and L.Lusanna, {\it The Einstein-Maxwell-Particle
System in the York Canonical Basis of ADM Tetrad Gravity: I) The
Equations of Motion in Arbitrary Schwinger Time Gauges.}, 2009
(arXiv 0907.4087).



\bibitem{13}L.Lusanna, {\it The Rest-Frame Instant Form of Metric Gravity},
Gen.Rel.Grav. {\bf 33}, 1579 (2001)(gr-qc/0101048).\hfill\break
 L.Lusanna and S.Russo, {\it A New Parametrization for Tetrad Gravity},
Gen.Rel.Grav. {\bf 34}, 189 (2002)(gr-qc/0102074).\hfill\break
R.DePietri, L.Lusanna, L.Martucci and S.Russo, {\it Dirac's
Observables for the Rest-Frame Instant Form of Tetrad Gravity in a
Completely Fixed 3-Orthogonal Gauge}, Gen.Rel.Grav. {\bf 34}, 877
(2002) (gr-qc/0105084).\hfill\break
 J.Agresti, R.De Pietri, L.Lusanna and L.Martucci, {\it
Hamiltonian Linearization of the Rest-Frame Instant Form of Tetrad
Gravity in a Completely Fixed 3-Orthogonal Gauge: a Radiation Gauge
for Background-Independent Gravitational Waves in a Post-Minkowskian
Einstein Spacetime}, Gen.Rel.Grav. {\bf 36}, 1055 (2004)
(gr-qc/0302084).\hfill\break
 J.Agresti, R.De Pietri, L.Lusanna and
L.Martucci, {\it Hamiltonian Linearization of the Rest-Frame Instant
Form of Tetrad Gravity in a Completely Fixed 3-Orthogonal Gauge: a
Radiation Gauge for Background-Independent Gravitational Waves in a
Post-Minkowskian Einstein Spacetime}, Gen.Rel.Grav. {\bf 36}, 1055
(2004) (gr-qc/0302084).

\bibitem{b}T.Buchert, {\it Dark Energy from Structure: a Status Report} (0707.2153).






\end{thebibliography}
\end{document}